\documentclass[11pt,a4paper]{article}
\pdfoutput=1 
\usepackage{jheppub} 
\usepackage{graphicx}
\usepackage{bm}

\usepackage{caption}
\usepackage{framed}
\usepackage{xcolor}
\usepackage{rotating}
\usepackage{empheq}
\usepackage{afterpage}

\usepackage[utf8]{inputenc}
\title{\boldmath Renormalons in quantum mechanics}

\author[a]{Cihan Pazarba{\c{s}}\i,}
\author[a,b]{Dieter Van den Bleeken}
\affiliation[a]{Physics Department, Bo\u{g}azi\c{c}i University\\
 34342 Bebek / Istanbul, Turkey}
\affiliation[b]{Secondary address\\
Institute for Theoretical Physics, KU Leuven\\
3001 Leuven, Belgium}

\emailAdd{cihan.pazarbasi@boun.edu.tr}
\emailAdd{dieter.van@boun.edu.tr}

\abstract{We present a nonrelativistic one-particle quantum mechanics whose perturbative S-matrix exhibits a renormalon divergence that we explicitely compute. The potential of our model is the sum of the 2d Dirac $\delta$-potential -- known to require renormalization -- and a 1d Dirac $\delta$-potential tilted at an angle. We argue that renormalons are not specific to this example and exist for a much wider class of potentials. The ambiguity in the Borel summation of the perturbative series due to the renormalon pole is resolved by the physical condition of causality through careful consideration of the $i\epsilon$ prescription. The suitably summed perturbative result coincides with the exact answer obtained through the operator formalism for scattering.}

\def\calo{{\cal O}}

\def\a{\alpha}
\def\b{\beta}

\def\f{\phi}

\def\f{{\mathrm{f}}}
\def\ini{{\mathrm{i}}}

\def\pv{{\mathbf{p}}}
\def\uv{{\mathbf{u}}}

\newcommand{\lbar}{\lower0.2ex\hbox{$\mathchar'26$}\mkern-10mu \lambda}

\begin{document}
    \maketitle
    \flushbottom
\section{Introduction}
Renormalons are divergent contributions to the perturbative series due to diagrams involving momentum integration of logarithms originating in renormalization \cite{Beneke:1998ui}. It is commonly assumed that this is a feature specific to relativistic quantum field theory (QFT) but we will show, by explicit construction, that this phenomenon also appears in nonrelativistic one-particle\footnote{Recently in \cite{Marino:2019wra, Marino:2019fuy} indications for renormalons in many body non-relativistic quantum mechanics were found.} quantum mechanics (QM). 

The relation between the divergent growth of perturbation theory and non-perturbative physics is an old subject that has received renewed interest in the context of resurgence, see \cite{LeGuillou:1990nq} for an overview of the early literature and \cite{Marino:2012zq,Aniceto:2018bis, Dunne:2015eaa} for more recent reviews. The study of perturbative series and their non-perturbative completion is relevant for and possibly fundamental to our understanding of 4d QFT, but often insight has been gained by studying simpler examples in quantum mechanics \cite{Bender:1969si, Bogomolny:1980ur, ZinnJustin:1981dx, Dunne:2014bca}. So far a one-particle quantum mechanics example with renormalons has not been considered\footnote{A notable exeption is \cite{Penin:1996zk} where a renormalon-like divergence in quantum mechanics was studied, but in an observable -- a scattering wave-packet -- that does not seem to have a standard QFT analog.} in the literature and this paper is a first step towards filling this gap. We hope that bringing the renormalon into the simpler and rigorously defined context of QM can be a step towards a further understanding of renormalons and the associated non-perturbative effects in QFT\footnote{Note that renormalons have been considered in various lower dimensional, simpler QFT models. See for example \cite{David:1983gz,Novikov:1984rf} for some of the earliest work, where the relation to the non-perturbative OPE was made explicit.}. 

Due to the simplicity of our model we are able to rigorously show -- by explicit calculation -- the existence of a renormalon divergence of the perturbative series of its S-matrix. This is important, as for 4d field theories it has so far been impossible to exclude a cancellation between various renormalon diagrams. Interestingly we will see that in our model indeed some cancellations take place, but a total non-zero contribution remains. This reflects itself in a growth $\propto(n-3)!$ in the order $n$ of perturbation theory, rather than the naively expected $\propto (n-1)!$. Additionally we use the formal tools of quantum mechanical scattering theory to compare the diverging perturbative series to the exact non-perturbative result. This reveals that Borel summation, using the correct prescription to evade poles, does indeed reconstruct the correct answer including the non-perturbative contribution. The nature of this non-perturbative effect, for example a possible semi-classical realization, would be interesting to further investigate in the future.

We start our paper by a short review of renormalons and their place in the theory of divergent series in section \ref{rensec}. In section \ref{2dd} we recall the quantum mechanics of the 2d $\delta$-potential and its renormalization, which is well-established but maybe not as well-known.  We then continue in section \ref{qmrend} by presenting the computation of a simple renormalon diagram in quantum mechanics. We focus on a simple example based on coupling the 2d $\delta$-potential to a 1d $\delta$ potential supported along a third direction. The main results of our paper are in section \ref{main}. We consider there a potential of the form $V=\lambda_0 \delta(x)\delta(y)+\kappa V_*$. The physical quantity we study is $\frac{1}{2}\left.\frac{\partial^2}{\partial\kappa^2}S(\pv_\f,\pv_\ini;\lambda,\kappa)\right|_{\kappa=0}$, i.e. the S-matrix exact in the renormalized coupling $\lambda$ and second order in $\kappa$. We'll discuss under which conditions on $V_*$ we expect renormalons to appear and work out in detail the case $V_*=\delta(\cos\theta z-\sin\theta y)$. The angle $\theta$ allows us to interpolate between the case $\theta=0$, where the model factorizes and the renormalon contributions are forced to cancel out among themselves, and the case $0<\theta<\frac{\pi}{2}$ where non-trivial interaction takes place and a non-zero renormalon contribution remains once all diagrams at a given order are summed. We compute the leading growth of the series coefficients due to the total renormalon contribution and discuss how this leads to a pole on the real axis in the Borel plane, resulting in a summation ambiguity. Alternatively one can sum the diagrams before performing the outer-loop momentum integral, which reproduces the same ambiguity. We point out that in this second summation procedure the ambiguity is naturally resolved by re-introducing the Feynman $i\epsilon$ prescription. This illustrates that in this case the summation ambiguity orginates from the limits $\epsilon\rightarrow 0$ and $n\rightarrow \infty$ not commuting. The link between renormalons and the $i\epsilon$ prescription was already suggested in some of the earliest studies on renormalons \cite{Olesen:1977ih}. In the Borel plane this corresponds to a deformation of the integration contour {\it below} the renormalon pole. In section \ref{exact} we use the operator formalism and exact knowledge of the Green's operator/resolvent of the 2d $\delta$-potential to rederive the perturbative result without exanding in the coupling $\lambda$. This calculation confirms the absence of further non-perturbative effects that could potentially have cancelled the non-perturbative contribution to the imaginary part due to the summation prescription. The exact calculation also makes the role of the $i\epsilon$ prescription fully transparent.

\section{Lightning review of renormalons and divergent series}\label{rensec}
\begin{figure}
	\begin{center}
	\includegraphics[scale=0.7]{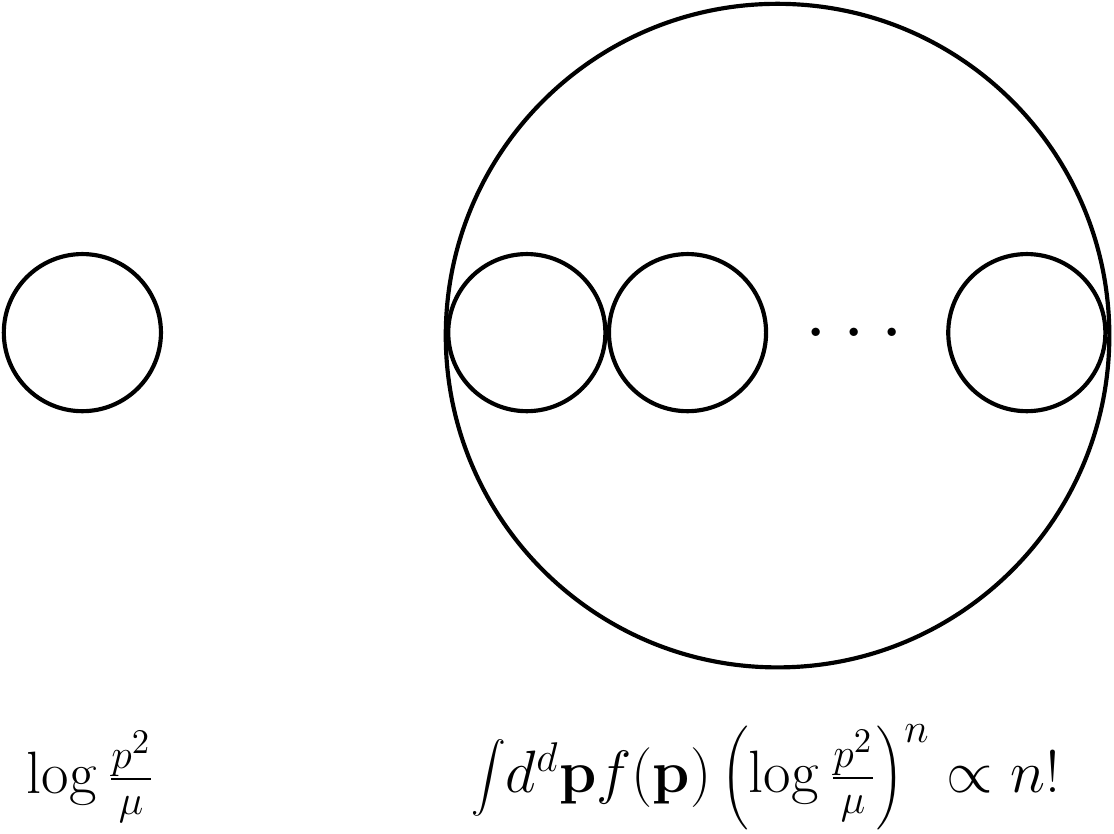}
	\end{center}
\caption{Left: A renormalized 1-loop diagram. Right: $n$ renormalized 1-loop diagrams inside a larger loop.}
\label{rendiaggen}
\end{figure}

The perturbative series of a physical quantity in a coupling constant $\lambda$ is often divergent rather than convergent. This happens when an instability appears under change of the phase of the complexified coupling constant, ruling out an analytic series expansion of this quantity in that coupling constant \cite{Dyson:1952tj}. In practice this typically manifests itself in a factorial growth of the series coefficients. Often the origin of this factorial growth is simply the combinatorial growth of the number of contributing diagrams as the order increases. But it can happen that a single diagram contributing at order $\lambda^n$ has size $\propto n!$\,. Indeed this is almost automatic in theories where renormalization leads to diagrams with logarithmic momentum dependence \cite{Gross:1974jv,Lautrup:1977hs,tHooft:1977xjm}, hence the name renormalon divergence. There are a number of excellent reviews \cite{LeGuillou:1990nq,Beneke:1998ui, ZinnJustin:2002ru, Beneke:2000kc,Shifman:2013uka} with explicit examples in QCD, QED and $\phi^4_4$, so it will suffice here to simply sketch how this comes about. Given a diagram that depends logarithmically on the momentum -- say a one-loop diagram after renormalization -- the theory will often contain higher order contributions made of $n$ consecutive insertions of this logarithmic diagram inside a larger loop -- see figure \ref{rendiaggen}-- leading to an integral of the form
\begin{equation}
I_n=\int d^d\pv\, f(\pv) \left(\log \frac{p^2}{\mu}\right)^n \label{renint}
\end{equation}
When $n$ is large this integral will be dominated where the logarithm is large, i.e. at large or small momentum. When $f(\pv)\sim p^{a}$ in the limit of large or small momentum a saddle point evaluation of the integral \eqref{renint} leads to respectively
\begin{equation}
I_n\propto \left(\pm\frac{a+d}{2}\right)^{n+1}n!
\end{equation}
Diagrams as on the right of figure \ref{rendiaggen} that have factorial contributions from large or small momentum regions are referred to as UV/IR renormalon diagrams respectively.

\begin{figure}
	\begin{center}
		\includegraphics[scale=0.7]{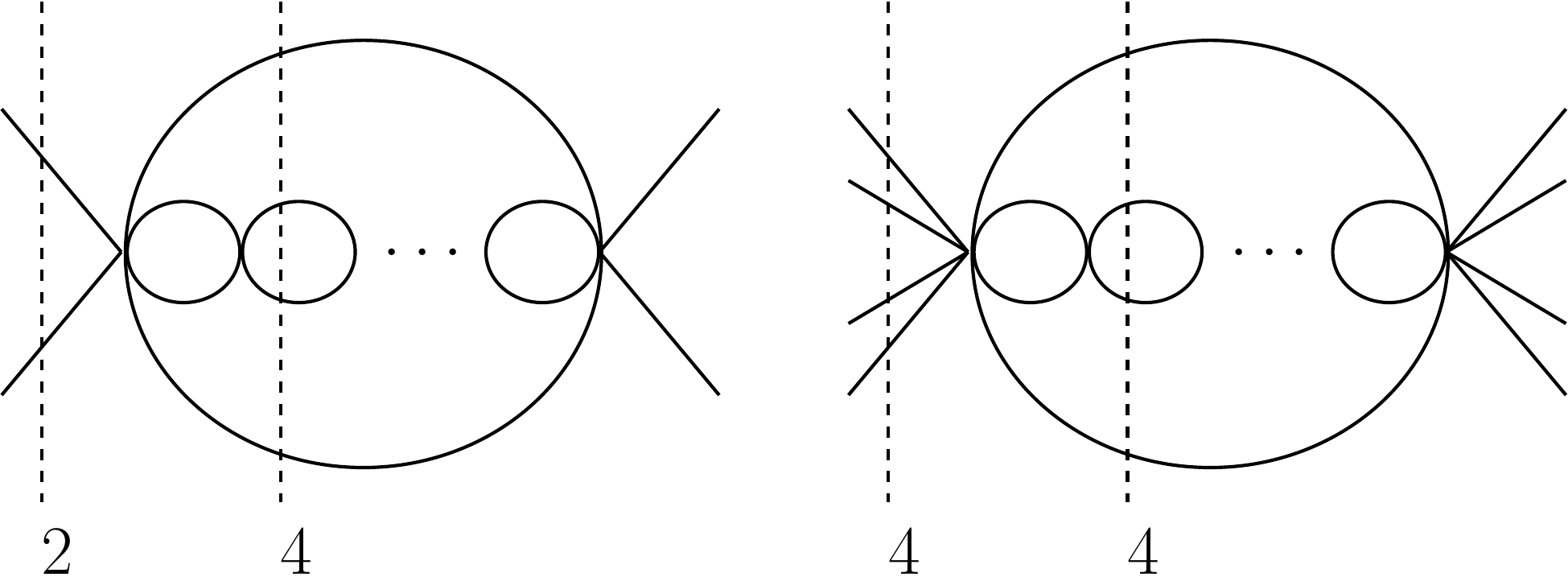}
	\end{center}
	\caption{Left: Renormalon type diagrams in 2-particle scattering violate particle number conservation.  Right: In 4-particle scattering particle number can be conserved.}
	\label{qmpart}
\end{figure}

Renormalization leading to a logarithmic momentum dependence is often associated with QFT but also exists in QM. A pedagogic dicussion of how this happens for the 2$\delta$-potential can be found in \cite{Jackiw:1995be} and we will review some aspects in section \ref{2dd}. Still this model, or at least its 2 particle S-matrix, does not exihibit renormalons. In QFT diagrams like the one on the right of figure \ref{rendiaggen} can appear in 2 particle scattering-- by attaching two external legs both left and right --  but this is not the case in QM as such a diagram would violate particle number conservation, see the left of figure \ref{qmpart}. But from this limitation it is at the same time clear that it can be evaded by attaching 4 external legs on both sides of the diagram of figure \ref{rendiaggen} and consider the 4 particle scattering matrix -- with a potential generating interaction between all four particles -- so that particle number is indeed conserved, see the right of figure \ref{qmpart}. This illustrates that quantum mechanics has all the ingredients for renormalons, at least if one chooses a potential that gets renormalized and that is interactive enough to allow for non-trivial multi-particle scattering. One can further simplify the setup considering all particles to have the same mass and interpreting the coordinates of the additional particles as extra spatial dimensions associated to a single particle. This brings us then to a model where -- instead of multi-particle scattering -- a single particle scatters of a background potential that has a part -- say the 2d $\delta$-potential -- that gets renormalized and an additional part that couples it to a third direction. That this idea is correct and that such models do really have a non-vanishing renormalon divergence in their perturbative S-matrix is what we will show in sections \ref{qmrend} and \ref{main}. It could be very interesting to look for renormalons in other QM observables of similar models but we leave this for future work.

Let us conclude this section with recalling some results on the connection between factorially diverging perturbative series --due to combinatorics or renormalon diagrams -- and non-perturbative effects, see e.g. \cite{Marino:2012zq} for an introduction. Given such a series
\begin{equation}
\langle \calo \rangle\sim \sum_{n=0}^\infty a_n \lambda^n\qquad a_n\sim A^{-n}(n-k)!\label{growth}
\end{equation}
one can use it to construct a finite quantity $\langle \calo \rangle$ through Borel summation:
\begin{equation}
\langle \calo \rangle=\sum_{n=0}^{k-1}a_n\lambda^n+\int_0^\infty ds\, e^{-\frac{s}{\lambda}}\sum_{n=k}^\infty \frac{a_n}{(n-k)!} s^{n-k}\label{borelresum}
\end{equation}
This summation procedure transforms the divergence of the perturbative series into non-perturbative contributions of size $e^{-\frac{A}{\lambda}}$. The key point is that the growth of the perturbative series is connected to the size of non-perturbative effects, both being characterized by the same parameter $A$. Note that the series in \eqref{borelresum} -- the Borel-transform -- is convergent but will have a pole at $s=A$. This is problematic when $A$ is positive real as it makes the integral \eqref{borelresum} a priory ill-defined. This can be remedied by redefining the integral via a contour just above or below the real axis, but at the cost of an ambiguity since these two choices lead to different answers.  The difference -- which corresponds to a discontinuity of $\langle\calo\rangle$ as a function of $\lambda$ -- is proportional to the residue of the integrand at the pole. The ambiguity is half this discontinuity with the sign $-/+$ depending on deformation above/below the pole. For growth of the form \eqref{growth} the ambiguity\footnote{As \eqref{growth} only provides the leading behaviour for large $n$ it provides information only about the pole closest to the origin of the Borel plane. Contributions to $n$ that are subleading but still divergent can lead to singularities further away from the origin. This means \eqref{bamb} is only the leading ambiguity, i.e. the full ambiguity can contain parts that vanishes faster than \eqref{bamb} in the limit $\lambda\rightarrow 0$.} is
\begin{equation}
\mathrm{amb}\langle \calo \rangle=\mp \pi i \left(\frac{\lambda}{A}\right)^{k-1} e^{-\frac{A}{\lambda}}\label{bamb}
\end{equation}

There are two types of resolution of this summation ambiguity. Sometimes a non-vanishing contribution from the singularity is physically required and in this case some additional non-perturbative physical information is needed to decide on the sign in \eqref{bamb} and resolve the ambiguity. In other situations there should be no extra contribution from the pole on physical grounds and the ambiguity gets canceled by an identical ambiguity of a further non-perturbative correction. This cancelation of ambiguities is related to the extension of the perturbative series to a trans-series and the theory of resurgence \cite{Marino:2012zq,Dunne:2015eaa,Aniceto:2018bis}. In QFT discussions of renormalons the observables considered are typically (Euclidean) Green's functions and the ambiguity they introduce can be removed by an OPE analysis \cite{David:1983gz,Novikov:1984rf}, see \cite{Beneke:1998ui,Shifman:2013uka} for reviews. Recently a further connection to transeries and resurgence has been proposed \cite{Maiezza:2019dht} The relation between renormalons and non-perturbative/power corrections has many phenomenological applications in QCD \cite{Beneke:2000kc}. In this paper we consider the (Lorentzian) S-matrix and the renormalon divergence is of the first type, namely the extra contribution \eqref{bamb} from the pole in the Borel plane will not be cancelled by other non-perturbative contributions, but will remain and its sign fixed by the $i\epsilon$ prescription which is equivalent to the physical condition of causality of scattering.

Since divergent perturbative series require non-perturbative contributions to complete them one is lead to ask if these non-perturbative effects can be independently understood. In the case of combinatorially driven growth this is the case \cite{Lipatov:1976ny} and the associated non-perturbative effects are instantons, saddle points in the semi-classical evaluation of the path integral. For renormalons such an independent interpretation is not universally established, although recently a large effort towards settling this question has been made \cite{Argyres:2012ka,Argyres:2012vv,Dunne:2012ae,Dunne:2012zk,Dabrowski:2013kba,Cherman:2013yfa,Anber:2014sda,Dunne:2015ywa,Fujimori:2018kqp}. We leave this issue in our model as an interesting open question, noting that the simple and mathematically rigorous setting of quantum mechanics should allow a precise answer to be formulated.

\section{Quantum mechanics with a 2d $\delta$-potential}\label{2dd}
In this section we review some aspects of quantum mechanics with a 2d $\delta$-potential, see e.g. \cite{Gosdzinsky:1990vz,Manuel:1993it,Jackiw:1995be}. As we will discuss this model requires renormalization, has a non-trivial -- but 1-loop exact -- $\beta$-function and a renormalization invariant energy scale $\Lambda=\mu e^{\frac{4\pi}{\lambda}}$ which is the energy of a non-perturbative bound state $E_\mathrm{b}=-\Lambda$.  What makes this model extra appealing is that the perturbative renormalization matches perfectly with a non-perturbative definition through the method of self-adjoint extension \cite{Albeverio:88}, as we will shortly recall at the end of this section. Since in the next sections we will couple this model to an additional third direction we will from the beginning discuss it in a 3 dimensional context but -- at least in this section -- the third direction trivially factorizes\footnote{More precisely $S_{\mathrm{3d}}(\pv_\f,\pv_\ini)=2\pi\delta(q_\f-q_\ini)S_{\mathrm{2d}}(\uv_\f,\uv_\ini)$. The reader interested only in a review of the 2d $\delta$ interaction can simply ignore all prefactors $2\pi\delta(q_\f-q_\ini)$ in this section.}. The starting point is the Hamiltonian (see appendix \ref{not} for our position and momentum space notation and conventions)
\begin{equation}
H=p^2+\lambda_0V_\star({\bf x})\qquad V_\star({\bf x})=\delta(x)\delta(y)\label{deltahamiltonian}
\end{equation}
We will proceed in a rather pedestrian way with the presentation reflecting a QFT treatment. Our aim is to compute the S-matrix of the model \eqref{deltahamiltonian}, describing the scattering of the particle off the background potential $V_\star$. It is standard practice to rewrite
\begin{equation}
S(\pv_\f,\pv_\ini)=\delta^3(\pv_\f-\pv_\ini)-2\pi i\, \delta(p_\f^2-p_\ini^2)\, t(\pv_\f,\pv_\ini)\label{Smatrix}
\end{equation}
The perturbative series for the on-shell $T$-matrix $t(\pv_\f,\pv_\ini)$ is the familiar Born-series and it is fully determined by the Fourier transform of the potential:
\begin{equation}
\hat V_\star(\pv)=2\pi\,\delta(q) 
\end{equation}
At $n^{\mathrm{th}}$ order in $\lambda_0$ there is a single diagram made of $n$ vertices connected by $n-1$ propagators  with the Feynmann rules
\begin{equation}
\star\ :\ \lambda_0\hat V_\star(\pv_{k-1}-\pv_k)\qquad
-\ :\ \int \frac{d^3\pv_k}{(2\pi)^3}\frac{1}{p_\f^2+i\epsilon-p_k^2}\label{FR1}
\end{equation}
More precisely we have ($\pv_0=\pv_\f, \pv_n=\pv_\ini$)
\begin{equation}
\star-\star-\ldots-\star\ = \ t^{(n)}(\pv_\f,\pv_\ini)=\lambda_0^n\int \left(\prod_{k=1}^{n-1}\frac{d^3\pv_k}{(2\pi)^3(p_\f^2+i\epsilon-p_k^2)}\right)\left(\prod_{k=1}^{n}\hat V_\star(\pv_{k-1}-\pv_k)\right)
\end{equation} 
The first order is the so-called Born-approximation, which in this case evaluates to
\begin{equation}
\star\ = \ t^{(1)}(\pv_\f,\pv_\ini)=2\pi\lambda_0\,\delta(q_\f-q_\ini)\label{naivezeroth}
\end{equation}

\subsection*{1-loop}

The need for renormalization in this model manifests itself at the next order -- which is the 1-loop order -- where a UV-divergent momentum integral appears:
\begin{equation}
\star-\star\ = \ t^{(2)}(\pv_\f,\pv_\ini)=2\pi\delta(q_\f-q_\ini)\, \frac{\lambda_0^2}{4\pi}\,\int_0^\infty\frac{du^2}{u_\f^2+i\epsilon-u^2}\label{naivefirst}
\end{equation}
One deals with this divergent integral in textbook fashion. First we regularize by introducing a UV momentum cutoff $\Omega$:
\begin{equation}
I_\Omega(z)=\frac{1}{4\pi}\int_0^\Omega\frac{du^2}{z-u^2}=\frac{1}{4\pi}\log z-\frac{1}{4\pi}\log\left(e^{i\pi}(\Omega-z)\right)
\end{equation}
To proceed we replace the bare coupling $\lambda_0$ by a physical coupling $\lambda$ at some fixed energy scale $\mu$ through $\lambda_0=\lambda+\lambda^2\Delta\left(\frac{\Omega}{\mu}\right)+\calo(\lambda^3)$. Observing that
\begin{equation}
\lambda_0+\lambda_0^2 I_\Omega(z)+\calo(\lambda_0^3)=\lambda+\lambda^2\left(\Delta\left(\frac{\Omega}{\mu}\right)+\frac{1}{4\pi}(\log z-\log(z-\Omega))\right)+\calo(\lambda^3)
\end{equation}
tells us to choose\footnote{Note that of course we could add an arbitrary (complex) constant to $\Delta$, however from \eqref{lphys} it follows that $\Delta\rightarrow \Delta+c$ can be absorbed by $\lambda\rightarrow \frac{\lambda}{1-c\lambda}$. The choice we make has the advantage that real $\lambda$ corresponds to a unitary S-matrix. This directly follows form \eqref{tfirst} and the fact that a unitary S-matrix requires the first order on-shell T-matrix to satisfy $t^{(1)}(\pv_\f,\pv_\ini)=t^{(1)}(\pv_\ini,\pv_\f)^*$. Of course this requirement only fixed $c$ to be real, but non-zero real $c$ amounts only to a rescaling of the momentum scale $\mu$, which is arbitrary in any case.}
\begin{equation}
\Delta(z)=\frac{1}{4\pi}\log z
\end{equation}
so that
\begin{equation}
\lim_{\Omega\rightarrow\infty}\left(\lambda_0+\lambda_0^2 I_\Omega(z)+\calo(\lambda_0^2)\right)=\lambda+\lambda^2\, l(z)+\calo(\lambda^3)
\end{equation}
where for future convenience we introduced the function
\begin{equation}
l(z)=\frac{1}{4\pi}\log \frac{e^{i\pi}z}{\mu}\label{ldef}
\end{equation}
The outcome of this renormalization procedure is to replace (\ref{naivezeroth}, \ref{naivefirst}) by 
\begin{eqnarray}
\star &\ =\ & t^{(1)}(\pv_\f,\pv_\ini)=2\pi\lambda\,\delta(q_\f-q_\ini)\label{tfirst}\\
\star-\star &\ =\ & t^{(2)}(\pv_\f,\pv_\ini)=2\pi\,\delta(q_\f-q_\ini)\, \lambda^2\,l(u^2_\f)
\end{eqnarray}
Imposing the result to be independent of the arbitrary scale $\mu$ leads to the 1-loop $\beta$-function
\begin{equation}
\beta(\lambda)=\frac{\lambda^2}{4\pi}+\calo(\lambda^3)\label{1loopb}
\end{equation}

\subsection*{All order}
This theory is so simple that the higher orders are easily analysed and can be directly summed. Indeed, note that
\begin{equation}
\star-\ldots-\star\ = \ t^{(n)}(\pv_\f,\pv_\ini)=2\pi\delta(q_\f-q_\ini)\, \lambda_0^n\left(\frac{1}{4\pi}\,\int_0^\infty\frac{du^2}{u_\f^2+i\epsilon-u^2}\right)^{n-1}
\end{equation}
This suggests that higher order renormalization simply amounts to repeating the 1-loop procedure via
\begin{equation}
\lambda_0\rightarrow \lambda\qquad\qquad \frac{1}{4\pi}\int_0^\infty\frac{du^2}{z-u^2}\rightarrow l(z)\label{renrules}
\end{equation}
It can be verified that this is indeed equivalent to the all order definition of the physical coupling
\begin{equation}
\lambda_0=\sum_{n=1}^\infty\left[\Delta\left(\frac{\Omega}{\mu}\right)\right]^{n-1}\lambda^n=\frac{\lambda}{1-\Delta\left(\frac{\Omega}{\mu}\right)\lambda}\label{lphys}
\end{equation}
In summary, after renormalization one finds
\begin{equation}
\star-\ldots-\star\ = \ t^{(n)}(\pv_\f,\pv_\ini)=2\pi\,\delta(q_\f-q_\ini)\, \lambda^n\,l(u_\f^2)^{n-1}\label{1ddn}
\end{equation}
So all order perturbation theory takes the simple form of a geometric series and the total answer is thus
\begin{equation}
t(\pv_\f,\pv_\ini)=\sum_{n=1}^\infty t^{(n)}(\pv_\f,\pv_\ini)= 2\pi\,\delta(q_\f-q_\ini)\,t_\star(u_\f^2)=\frac{2\pi\,\delta(q_\f-q_\ini)\,\lambda}{1-\frac{\lambda}{4\pi}(\log \frac{u^2_\f}{\mu}+i\pi)}\label{2dt}
\end{equation}
where for later convenience we separately define
\begin{equation}
t_\star(z)=\sum_{n=0}^\infty l(z)^{n-1}\lambda^n=\frac{\lambda}{1-\frac{\lambda}{4\pi}\log\frac{e^{i\pi} z}{\mu}}\label{tstar}
\end{equation}

\subsection*{Discussion}
Let us now interpret the results of the calculation performed above. Via the all order definition of the physical coupling \eqref{lphys} one can compute the all order $\beta$-function
\begin{equation}
\beta(\lambda)=\frac{\lambda^2}{4\pi}\label{beta}
\end{equation}
It is interesting to note that this coincides with \eqref{1loopb}, implying the $\beta$-function is one-loop exact. From \eqref{beta} one computes the running coupling
\begin{equation}
\bar\lambda(p^2)=\frac{\lambda}{1-\frac{\lambda}{4\pi}\log \frac{p^2}{\mu}}=\frac{4\pi}{\log\frac{\Lambda}{p^2}}\label{runcoup}
\end{equation}
which reveals a renormalization invariant scale
\begin{equation}
\Lambda=\mu\, e^{\frac{4\pi}{\lambda}}\label{renscal}
\end{equation}
In accord with renormalization theory the dependence of \eqref{2dt} on the energy scale goes purely through the running coupling:
\begin{equation}
t(\pv_\f,\pv_\ini)=2\pi \delta(q_\f-q_\ini)\frac{\bar\lambda(u_\f^2)}{1-\frac{i}{4}\bar\lambda(u_\f^2)}\label{runningt}
\end{equation}
Interestingly this theory is both UV and IR free without being trivial. Although the running coupling \eqref{runcoup} has a Landau pole at energy $\Lambda$, the S-matrix (\ref{Smatrix}, \ref{runningt}) remains perfectly finite at this energy. It does have a pole however at {\it negative} energy $E_\mathrm{b}=-\Lambda$, showing that the proper physical interpretation of $\Lambda$ is that of the energy of a non-perturbative bound state. Interestingly -- and contrary to the 1d $\delta$-potential -- the bound state exists both for positive and negative $\lambda$. Let us stress -- as this is important for the correct interpetation of the following sections -- that the model is well defined if and only if $\lambda$ is real, or equivalently for all positive real values of $\Lambda$.

Surprisingly one is able to extract non-perturbative information of the model -- the bound state energy -- through a purely perturbative calculation enhanced with renormalization. Note that so far the perturbative series considered was perfectly convergent so the non-pertrubative boundstate is not connected to any divergence\footnote{Note that we focussed here simply on the S-matrix. It is not excluded that the non-perturbative boundstate can be linked to the divergence of the perturbative series of another observable and it would be interesting to investigate this.}. In the next sections we will see that it can be linked to the divergence of the perturbative S-matrix once we couple the particle to an additional potential.

\subsection*{Exact solution}
Although the non-perturbative bound state emerged out of the perturbative treatment above one might wonder if the S-matrix could not get additional non-perturbative contributions that are missed perturbatively. Due to the simplicity of the 2d $\delta$-model one can actually solve it exactly which not only confirms the renormalized perturbative calculations above but additionally shows that that answer is complete. The advantage of QM compared to QFT is that we have an explicit non-perturbative definition provided by the Schr\"odinger equation with a self-adjoint Hamiltonian. Although they will not be applied in the remainder of of the paper, we shortly mention the results obtained by treating the model through the method of self-adjoint extensions as they are quite beautiful and put the work in this and the following sections on firmer footing. For further details including a more precise mathematical treatment see \cite{Albeverio:88}.

The idea is to replace \eqref{deltahamiltonian} -- which is a Hamiltonian defined on all of $\mathbb{R}^3$ {}\footnote{The discussion \cite{Albeverio:88} is on $\mathbb{R}^2$, but again we add a trivial $3^{\mathrm{th}}$ direction as it makes comparison to the previous part of this section more straightforward.} -- by the free Hamiltonian on $\mathbb{R}^3\backslash\mathbb{R}$ -- the line being removed is the origin of the $xy$-plane -- supplemented by a boundary condition at $x=y=0$. The condition that the 'free' Hamiltonian $H=p^2$ be self-adjoint with respect to this boundary condition strongly restricts the options, so much so that all posibilities can be classified. Although a priory this could lead to point-interactions which are not described by a $\delta$-potential -- as indeed in general it does -- this is not the case in this setting. To be precise let us decompose the wavefunction as ($x=r\cos\phi, y=r\sin\phi$)
\begin{equation}
\psi(x,y,z)=\int_{-\infty}^\infty \frac{dq}{2\pi} \sum_{m=-\infty}^\infty \psi_m(r,q)e^{i(m\phi+qz)}
\end{equation} For $m\neq 0$ the only allowed boundary condition is simply that the wave-function remain finite as $r\rightarrow 0$, but the boundary condition on $\psi_0(r,q)$ can be non-trivial. Those boundary conditions that lead to a self-adjoint Hamiltonian are parameterized by a positive real parameter $\Lambda$ and read
\begin{equation}
\lim_{r\rightarrow 0} \left(\frac{\psi_0(r,q)}{r\partial_r\psi_0(r,q)}-\log \frac{\sqrt{\Lambda} r}{2}\right)=\gamma
\end{equation}
where $\gamma$ is the Euler-Mascheroni constant. One can then solve the time-independent Schr\"odinger equation with these boundary conditions to find scattering states and a bound state:
\begin{eqnarray}
\psi_{0,u,q}(r,q')&=&\frac{2\pi\delta(q-q')\sqrt{u}}{|i\pi+\log\frac{\Lambda}{u^2}|}\left(\pi Y_0(ur)+\log\frac{\Lambda}{u^2}\,J_0(ur)\right)\label{scat1}\\
\psi_{m,u,q}(r,q')&=&2\pi\delta(q-q')\sqrt{u}\,J_{m}(ur)\qquad\qquad m\neq 0\label{scat2}\\
\psi_{\mathrm{b}}(r,q)&=&2\pi\delta(q)\sqrt{\frac{\Lambda}{\pi}}K_0(\sqrt{\Lambda}r)
\end{eqnarray}
The scattering states have energy $E_{m,u,q}=u^2+q^2=p^2$ while the bound-state has energy $E_\mathrm{b}=-\Lambda$. Matching the bound state energy with the perturbative calculation above allows to identify the parameter $\Lambda$ of the self-adjoint extension with the renormalization invariant scale \eqref{renscal}. The non-trivial test is then to compare the scattering amplitude as defined by the scattering states (\ref{scat1},\ref{scat2}) with the perturbative on-shell T-matrix \eqref{runningt}. A short calculation -- since only $m=0$ leads to non-trivial scattering -- provides perfect agreement.

\section{A renormalon diagram in quantum mechanics}\label{qmrend}
This section discusses a first diagram for an example potential. In the next section we will discuss the totality of all diagrams, both for more general potentials as well as the example considered here. 

\begin{figure}
	\begin{center}
		\includegraphics[scale=0.7]{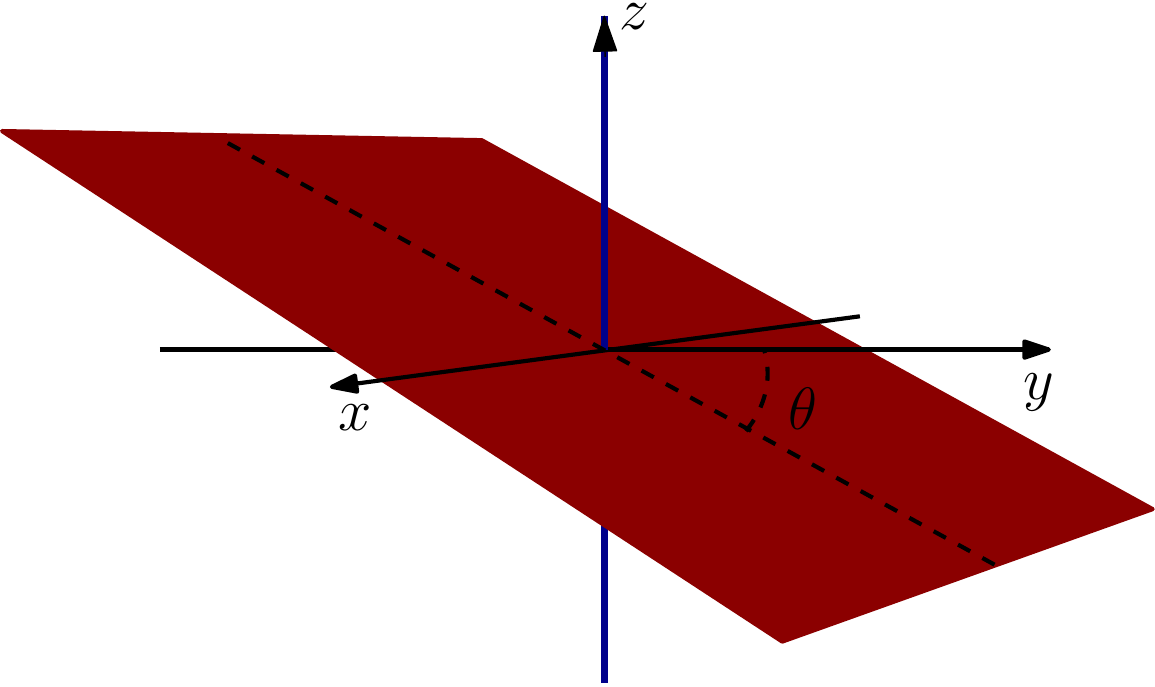}
	\end{center}
	\caption{Support of our example potential. The blue line -- coinciding with the $z$-axis -- corresponds to $V_\star=\delta(x)\delta(y)$, while the red plane corresponds to\\ $V_*=\delta(\cos\theta\,z-\sin\theta\,y)$.}
	\label{potsup}
\end{figure}

With the aim to generate a renormalon and motivated by the discussion of section \ref{rensec} we add to the 2d $\delta$-potential an extra potential that couples non-trivially to the 3$^{\mathrm{th}}$ direction:
\begin{equation}
H=p^2+\lambda_0V_\star+\kappa V_*\label{model}
\end{equation}
A simple choice is to take as the extra piece a 1d $\delta$ potential. To make sure the theory does not simply factorize we put the support of the 1d $\delta$ at an angle to the $xy$-plane:
\begin{equation}
V_*=\delta(\cos\theta\, z-\sin\theta\, y)\label{1ddpot}
\end{equation}
see also figure \ref{potsup}.
Keeping the angle $\theta$ a free parameter will allow a check on our results since through the limit $\theta\rightarrow 0$ we can compare to the case where the S-matrix factorizes:
\begin{equation}
S_{\theta=0}(\pv_\f,\pv_\ini)=S_{\mathrm{1d}\,\delta}(q_\f,q_\ini)S_{\mathrm{2d}\,\delta}(u_\f,u_\ini)\,.
\end{equation}
By adding a new part to the potential and introducing a second coupling $\kappa$ we introduce a whole new set of diagrams to the calculation of the perturbative S-matrix. For our discussion -- in this and the following sections -- it will be sufficient to focus only on diagrams quadratic in $\kappa$. Formally we could say that the observable of our interest is $ \frac{1}{2}\left.\frac{\partial^2}{\partial\kappa^2}S(\pv_\f,\pv_\ini;\lambda,\kappa)\right|_{\kappa=0}$, in practice it means we will work to all order in $\lambda$ and at second order in $\kappa$.

\begin{figure}
	\begin{center}
		\includegraphics[scale=0.7]{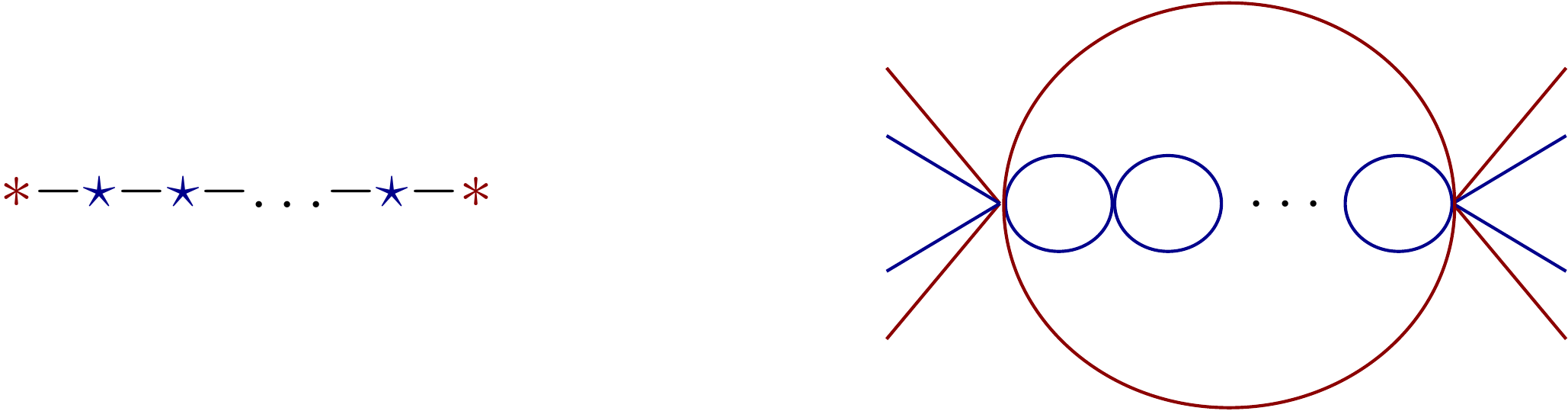}
	\end{center}
	\caption{A renormalon-type diagram. On the left the diagram describing one-particle scattering off a potential. On the right a corresponding diagram in the language of 4-particle scattering. }
	\label{potvspart}
\end{figure}

In this two parameter perturbation theory there are two types of vertices: $\star$ and $*$. The Feynmann rules \eqref{FR1} get extended by
\begin{equation}
*\ :\ \kappa\,\hat V_*(\pv_{k-1}-\pv_{k})\label{FR2}
\end{equation}
Given the $(\log u^2)^n$ behaviour of the diagram $\star-\star-\ldots -\star$ established in \eqref{1ddn}, one expects a renormalon might appear once this diagram finds itself inside a bigger loop. This can be done by squeezing the diagram between two additional $*$ vertices -- see also figure \ref{potvspart} -- and explains why it is the second order in $\kappa$ where we expect the phenomenon to first appear.

Let us now show that indeed this intuition is correct by explicit computation. For the example \eqref{1ddpot} one has ($\pv=(v,w,q)$)
\begin{equation}
*\ :\ \kappa\,(2\pi)^2 \delta(v_{k-1}-v_k)\delta\left(\cos\theta(w_{k-1}-w_{k})+\sin\theta(q_{k-1}-q_{k})\right) \label{1ddelfour}
\end{equation}
Applying the Feynmann rules, performing some integration and applying the renormalization \eqref{renrules} it follows that\footnote{See the next section for more details.}
\begin{equation}
*-\star-\star-\ldots -\star-*\ =\ \lambda^n \kappa^2\cos^2\theta \int \frac{dq}{2\pi}\frac{l(p_\f^2-q^2)^{n-1}}{\left((q_\f-q)(q+\tilde{q}_\f)+i\epsilon\right)\left((q_\ini-q)(q+\tilde{q}_\ini)+i\epsilon\right)}\label{1stint}
\end{equation}
where we used the shorthand
\begin{equation}
\tilde{q}_\alpha= \cos 2\theta\,q_\alpha-\sin 2\theta\, w_\alpha\qquad\qquad (\alpha=\f, \ini)\label{tildedef}
\end{equation}
We remind the reader that the function $l$ is essentially the logarithm, see \eqref{ldef}. The integral \eqref{1stint} is indeed of the generic renormalon type \eqref{renint}. The logarithm becomes large when $q^2\approx p_\f^2$ or $q^2\rightarrow \infty$. A careful analysis -- see appendix \ref{intap} -- reveals that factorial contributions to the integral \eqref{1stint} around $|q|=p_\f$ cancel each other while this is not the case at large momentum. Using that the rational part of the integrand in \eqref{1stint} decays like $q^{-4}=q^{-2\times\frac{3}{2}-1}$ for large $q$ and via formula \eqref{keyint} we find that for large $n$
\begin{equation}
\boxed{\ *-\star-\star-\ldots -\star-*\ \sim\  2\kappa^2\cos^2\theta\,\mu^{-\frac{3}{2}}\left(\frac{\lambda}{6\pi}\right)^n\,(n-1)!\ }\,.\label{1stres}
\end{equation}
where the factor $6\pi=\frac{3}{2}\times 4\pi$ is to be interpreted as a multiple of the (inverse of the) $\beta$-function coefficient \eqref{1loopb}. The appearance of $\frac{3}{2}$ is no coincidence and set by dimensional analysis. In 3d the on-shell T-matrix $t$ has dimension of length while $\kappa$ has dimension of inverse length. So the only way the other scale $\mu$ -- with dimension of inverse length squared -- can appear is with a power $-\frac{3}{2}$, fixing $\rho=\frac{3}{2}$ in \eqref{keyint}.

The result \eqref{1stres} is important in that it manifestly shows that also in non-relativistic 1-particle QM renormalon diagrams appear and that they lead to factorial growth through exactly the same mechanism as in QFT, i.e. integration in momentum space over an integrand that includes a high power of a logarithmic momentum dependence due to a large number of renormalized 1-loop diagrams inside a larger loop. Given the discussion in section \ref{rensec} we conclude there is a pole at $s=6\pi$ in the Borel plane when summing all the $*-\star-\star-\ldots -\star-*$ diagrams. When $\lambda$ is positive this will lead to an ambiguity \eqref{bamb} of the form
\begin{eqnarray}
\mathrm{amb}\left(\sum_n *-\star-\star-\ldots -\star-*\right)&=&\mp 2\pi i\,\kappa^2\, \cos^2\theta\,\mu^{-\frac{3}{2}}e^{-\frac{\lambda}{6\pi}}\\
&=&\mp 2\pi i\,\kappa^2\, \cos^2\theta\,\Lambda^{-\frac{3}{2}}\label{1stbamb}
\end{eqnarray}
It is interesting to note that while \eqref{1stres} appears not manifestly renormalization invariant, the corresponding ambiguity  \eqref{1stbamb} obtained from ressumation is manifestly renormalization invariant as it can be expressed purely in terms of the renormalization invariant scale $\Lambda$, defined in \eqref{renscal}.

Before one draws conclusions it should be realized that the diagrams considered above form only a subset of all the diagrams contributing to the S-matrix, and so one cannot directly extrapolate these results to the actual physical observable. Indeed, note that the growth \eqref{1stres} and the corresponding non-perturbative contribution \eqref{1stbamb} do {\it not} vanish as $\theta\rightarrow 0$. But because in that limit the full S-matrix is simply the product of the 1d and 2d $\delta$ S-matrices there should be no divergence nor an extra non-perturbative contribution. This indicates that there are further factorially growing sets of diagrams in the theory and that -- at least at $\theta=0$ -- these will cancel the growth \eqref{1stres}. This motivates us to carefully work through all diagrams in the next section, which will confirm such a cancellation at $\theta=0$ but will also show that when $\theta\neq 0$ the cancellation is not complete and a total factorial growth remains.

\section{Renormalons: all order perturbation theory} \label{main}
As illustrated in the last section, renormalon diagrams leading to factorial growth appear also in 1-particle QM. In this section we investigate this in more detail, carefully working out all diagrams for the model \eqref{model}. We will start with the potential $V_*$ arbitrary so we can understand more generally under which conditions renormalons can appear. We then specialize again to \eqref{1ddpot} to provide an explicit fully worked out example. After exhibiting the factorial growth we will consider the Borel summation and its ambiguity, show how it can be rephrased as an ambiguity of a momentum space integral and how that ambiguity is naturally resolved through the Feynman $i\epsilon$ prescription. An exact treatment in the next section confirms the perturbative results of this section.

\subsubsection*{First order in $\kappa$}
We'll analyze all diagramatic contributions to the on-shell T-matrix to arbitrary order in $\lambda$ and second order in $\kappa$, using the Feynman rules (\ref{FR1},  \ref{FR2}) together with the renormalization \eqref{renrules}. Although our interest is in the part of the S-matrix quadratic in $\kappa$ it will be useful to first consider the linear part, as some structures appearing there will have a role to play at second order. The first order consists of all diagrams with a single $*$ vertex, they can be easily listed and computed to be\footnote{The complex conjugate of the integral in \eqref{cceq} should be performed {\it without} changing the sign of the $i\epsilon$ term.}
\begin{eqnarray}
* &\ =\ &\kappa\hat V_*(\pv_\f-\pv_\ini)\\
\star-\ldots-\star-* &\ =\ & l(u_\f^2)^{n-1}\lambda^{n}\kappa\,I^{(1,1)}(q_\f,\pv_\ini)\\
*-\star-\ldots-\star &\ =\ & l(u_\f^2)^{n-1}\lambda^{n}\kappa\, I^{(1,1)}(q_\ini,\pv_\f)^*\label{cceq}\\ 
\underbrace{\star-\ldots-\star}_{n-a}-*-\underbrace{\star-\ldots-\star}_{a} &=& l(u_\f^2)^{n-a-1}\,l(u_\ini^2)^{a-1}\,\lambda^{n}\,\kappa\, I^{(1,2)}(q_\f,q_\ini)
\end{eqnarray}
Here two integrals appear:
\begin{eqnarray}
	I^{(1,1)}(\pv_\alpha,q_\beta)&=&\int \frac{d^2\uv}{(2\pi)^2}\frac{\hat V_*(\uv_\a-\uv,q_\a-q_\b)}{p_\f^2-q_\a^2+i\epsilon-u^2}\label{kint1}\\
	I^{(1,2)}(q_\alpha,q_\beta;z)&=&\int \frac{d^2\uv}{(2\pi)^2}\frac{d^2\uv'}{(2\pi)^2}\frac{\hat V_*(\uv-\uv',q_\a-q_\b)}{(p_\f^2-q_\alpha^2+i\epsilon-u^2)(p_\f^2-q_\beta^2+i\epsilon-u'^2)}\label{kint2}
\end{eqnarray}
In the particular example \eqref{1ddpot} these integrals evaluate to
\begin{eqnarray}
	I^{(1,1)}(\pv_\a,q_\b)&=&\frac{\cos\theta}{\cos^2\theta(p_\f^2-p^2_\a)+(q_\a-q_\b)(q_\b+\tilde q_\a)}\label{1ddint1}\\
	I^{(1,2)}(q_\a,q_\b)&=&\frac{1}{(2\pi)^2\cos\theta}\int \frac{dv dw}{(p_\f^2+i\epsilon-q_\a^2-v^2-(w-\tan\theta\, q_\a)^2)(p_\f^2+i\epsilon-q_\b^2-v^2-(w-\tan\theta\, q_\b)^2)}\nonumber\\
	&=&\frac{\cos\theta}{2\pi|q_\a^2-q_\b^2|F}\log\frac{2\sqrt{(p_\f^2-q_\a^2)(p_\f^2-q_\b^2)}}{2 p_\f^2-q_\a^2-q_\b^2+\sec^2\theta|q_\a^2-q_\b^2|F-\tan^2\theta (q_\a-q_\b)^2}\label{1ddint2}
\end{eqnarray}
where
\begin{equation}
F=\sqrt{1-4\sin^2\theta\,\frac{ q_1q_2+p_\f^2\cos^2\theta}{(q_1+q_2)^2}}
\end{equation}
see \eqref{tildedef} for the definition of $\tilde q$.

\subsubsection*{Second order in $\kappa$}
This is the order where we expect renormalon diagrams to appear. As a starting point for our discussion we present the expressions for all diagrams with two $*$ vertices and an arbitraty number $n$ of $\star$ vertices in table \ref{secondord}. 
\afterpage{\clearpage}
\begin{sidewaystable}[p]
	\begin{minipage}{22cm}
		
		\vspace{0.1cm}
		\begin{center}
			\caption{\label{secondord} Expressions for all 8 types of diagrams appearing at order $\kappa^2$. Only the diagrams in the box can lead to renormalons.}
		\end{center}
		\begin{eqnarray}
		*-* &\ =\ & \kappa^2\int \frac{dq}{2\pi}I^{(2,1)}(\pv_\f,\pv_\ini,q)\label{k2int1}\\
		\star-\ldots-\star-*-*&=& l(u_\f^2)^{n-1}\kappa^2\lambda^{n}\int \frac{dq}{2\pi}I^{(2,2)}(q_\f,\pv_\ini,q)\label{k2int2}\\
		*-*-\star-\ldots-\star&=&l(u_\ini^2)^{n-1}\kappa^2\lambda^{n}\int \frac{dq_1}{2\pi}I^{(2,2)}(q_\ini,\pv_\f,q)^*\label{k2int3}\\
		\underbrace{\star-\ldots\star}_{a}-*-*-\underbrace{\star-\ldots-\star}_{n-a}&=&l(u_\f^2)^{a-1}l(u_\ini^{2})^{n-a-1}\kappa^2\lambda^{n}\int\frac{dq}{2\pi}I^{(2,3)}(q_\f,q_\ini,q)\label{k2int4}
		\end{eqnarray}
		\begin{empheq}[box=\fbox]{align}
		*-\star-\ldots-\star-*&\ =\ \kappa^2 \lambda^{n}\int \frac{dq}{2\pi}I^{(1,1)}(\pv_\f,q)I^{(1,1)}(\pv_\ini,q)^*l(p_\f^2-q^2)^{n-1}\label{k2int5}\\
		\underbrace{\star-\ldots-\star}_{a}-*-\underbrace{\star-\ldots-\star}_{n-a}-*&\ =\  l(u_\f^2)^{a-1}\kappa^2\lambda^{n}\int\frac{dq}{2\pi}I^{(1,2)}(q_\f,q)I^{(1,1)}(\pv_\ini,q)^* l(p_f^2-q^2)^{n-a-1}\label{k2int6}\\
		*-\underbrace{\star-\ldots-\star}_{n-a}-*-\underbrace{\star-\ldots-\star}_{a}&\ =\  l(u_\ini^2)^{a-1}\kappa^2\lambda^{n}\int\frac{dq}{2\pi}I^{(1,2)}(q_\ini,q)^*I^{(1,1)}(\pv_\f,q) l(p_f^2-q^2)^{n-a-1} \label{k2int7}\\
		\underbrace{\star-\ldots-\star}_{a}-*-\underbrace{\star-\ldots-\star}_{n-a-b}-*-\underbrace{\star-\ldots-\star}_{b}&\ =\ l(u_\f^2)^{a-1}l(u_\ini^2)^{b-1}\kappa^2\lambda^{n}\int\frac{dq}{2\pi}I^{(1,2)}(q_\f,q)I^{(1,2)}(q_\ini,q)^*l(p_\f^2-q^2)^{n-a-b-1}\label{k2int8}
		\end{empheq}
		With -- in addition to (\ref{kint1}, \ref{kint2}) -- the definitions
		\begin{eqnarray}
		I^{(2,1)}(\pv_\a,\pv_\b,q)&=&\int \frac{d^2\uv}{(2\pi)^2}\frac{\hat V_*(\uv_\a-\uv,q_{\a}-q)\hat V_*(\uv-\uv_{\b},q-q_{\b})}{(p_\f^2-q^2+i\epsilon-u^2)}\\
		I^{(2,2)}(q_\a,\pv_\b,q)&=&\int \frac{d^2\uv_1}{(2\pi)^2}\frac{d^2\uv_2}{(2\pi)^2}\frac{\hat V_*(\uv_{1}-\uv_{2},q_{\a}-q)\hat V_*(\uv_{2}-\uv_\b,q-q_{\b})}{(p_\f^2-q^2_\a+i\epsilon-u_1^2)(p_\f^2-q^2+i\epsilon-u_2^2)}\\
		I^{(2,3)}(q_\a,q_\b,q)&=& \int\frac{d^2\uv_1}{(2\pi)^2}\frac{d^2\uv_2}{(2\pi)^2}\frac{d^2\uv_3}{(2\pi)^2} \frac{\hat V_*(\uv_{1}-\uv_{2},q_{\a}-q)\hat V_*(\uv_{2}-\uv_{3},q-q_{\b})}{(p_\f^2-q_\a^2+i\epsilon-u_1^2)(p_\f^2-q^2+i\epsilon-u_2^2)(p_\f^2-q_\beta^2+i\epsilon-u_3^2)}
		\end{eqnarray}
		
	\end{minipage}
\end{sidewaystable}

The first 4 types of diagrams (\ref{k2int1}-\ref{k2int4}) cannot grow factorially in $n$, since the integrands of the $q$ integral are $n$ independent. This implies we can safely ignore them at large $n$ and so we will not consider them further. The four remaining types (\ref{k2int5}-\ref{k2int8}) are what we are interested in. They contain an integral over $q$ of an integrand containing\footnote{We remind the definition \eqref{ldef}.} $(\log \frac{q^2}{\mu})^{n}$ and -- as in \eqref{renint} -- can lead to factorial growth in $n$. Indeed the diagrams \eqref{k2int5} are the ones we worked out in an example in the previous section confirming this factorial growth.

First let us point out that the loop integral in (\ref{k2int5}-\ref{k2int8}) containing the log's is over $q$, the momentum associated to the $z$-direction. Now observe that if we would choose the potential $V_*$ to be independent of this direction -- so that $\hat V_*\propto \delta(q)$ -- then this loop integral -- via (\ref{kint1}, \ref{kint2}) -- would become trivial and no factorial growth is generated. This shows that although the renormalization of the 2d $\delta$ potential $V_\star$ is crucial, so is the coupling to an additional potential that depends on a 3$^{\mathrm{rd}}$ direction. It appears to be the analog in one-particle mechanics of the need for more than 2 particles to generate renormalons in a multi-particle scattering setup, which -- as we argued in section \ref{rensec} -- is due to particle number conservation in QM (as opposed to QFT).

To understand if factorial growth does in fact appear one needs to be more precise about the non-logarithmic parts of the integrands of (\ref{k2int5}-\ref{k2int8}). These are formed by products of the first order integrals (\ref{kint1}, \ref{kint2}) which are in term determined through $\hat{V}_*$. Possible factorial growth for large $n$ originates in the momentum regions where the logarithm is large, either $q^2\approx p_\f^2$ or $q^2\gg \mu$. As we show in appendix \ref{intap} there is no net contribution from the first region while the contribution of the second region is fully determined by the large momentum behaviour of the non-logarithmic part of the integrand:
\begin{equation}
\int_{-\infty}^\infty \frac{dq}{2\pi}\, f(q)l(p_\f^2-q^2)^{n-1}\sim\frac{2\mu^{-\rho}}{(4\pi\alpha)^{n}} \,\left(\frac{n}{\rho}\right)^\sigma  \,(n-1)!\label{asform1}
\end{equation}
when for large $|q|$ the function $f(q)$ decays as
\begin{equation}
f(q)\sim |q|^{-2\rho-1}(\log q^2)^\sigma\label{asform2}
\end{equation}
The upshot is that the presence of renormalons is a feature of the large $|q_\beta|$ behaviour of the integrals (\ref{kint1}, \ref{kint2}) which in turn is fully determined by the large $|q|$ behaviour of the potential $\hat V_*(\uv, q)$. For this reason we expect the presence of renormalons to be a robust feature, not at all specific to the concrete example \eqref{1ddpot} that we will analyze in detail below. Indeed, if one would slightly change \eqref{1ddpot} in a way that that the second $\delta$-function in \eqref{1ddelfour} is replaced by another function -- say a Gaussian -- that is peaked around $|q|\approx |\uv|$ then this will not drastically change the large $|q|$ behaviour of the integrals (\ref{kint1}, \ref{kint2}) and one would expect renormalons to remain present. Understanding the precise mathematical conditions on $\hat V_*(\uv, q)$ for which a non-zero, non-cancelling set of renormalons appears would be interesting, but we leave it for future work.

\subsubsection*{Concrete example}
Let us now specialize to the specific example \eqref{1ddpot}, for which we computed the expressions (\ref{1ddint1}, \ref{1ddint2}). Their large $|q|$ decay is given by
\begin{eqnarray}
I^{(1,1)}(\pv_\a,q_\b)&=&- \cos\theta\, q_\b^{-2}+\calo(q_\b^{-3})\label{largeqI1}\\
I^{(1,2)}(q_\a,q_\b)&=&\frac{\cos\theta}{4\pi}q_\beta^{-2}\left(\log \frac{q_\b^2}{\mu}-4\pi\, l(p_\f^2-q_\a^2)+\log \cos^2\theta\right)+\calo(q_\b^{-3})\label{largeqI2}
\end{eqnarray}
Note that in the second line we also included a first subleading term as this will come to play a role.  Using these formulas and (\ref{asform1}, \ref{asform2}) it is straightforward to compute the growth of the 4 relevant diagrams (\ref{k2int5}-\ref{k2int8}):
\begin{align}
&*-\star-\ldots-\star-*\nonumber\\
&\ \sim C_n (n-1)!\,\left(1+\calo(n^{-1})\right)\qquad \qquad \qquad \qquad \qquad C_n=2\cos^2\theta \kappa^2 \mu^{-\frac{3}{2}}\left(\frac{\lambda}{6\pi}\right)^n\label{grint5}\\
&\underbrace{\star-\ldots-\star}_{a}-*-\underbrace{\star-\ldots-\star}_{n-a}-*\nonumber\\
&\ \sim \  -C_n\,l(u_\f^2)^{a-1}\left((n-a)!-(6\pi\, l(u_\f^2)+\frac{3}{2}\log\cos^2\theta)(n-a-1)!\right)\left(1+\calo(n^{-1})\right) \label{grint6}\\
&*-\underbrace{\star-\ldots-\star}_{n-a}-*-\underbrace{\star-\ldots-\star}_{a}\nonumber\\
&\ \sim \  -C_n\,l(u_\ini^2)^{a-1}\left((n-a)!-(6\pi\, l(u_\ini^2)+\frac{3}{2}\log\cos^2\theta)(n-a-1)!\right)\left(1+\calo(n^{-1})\right) \label{grint7}\\
&\underbrace{\star-\ldots-\star}_{a}-*-\underbrace{\star-\ldots-\star}_{n-a-b}-*-\underbrace{\star-\ldots-\star}_{b}\nonumber\\
&\ \sim\ C_n\,l(u_\f^2)^{a-1}l(u_\ini^2)^{b-1}\Big((n-a-b+1)!-(6\pi\, l(u_\ini^2)+6\pi\, l(u_\ini^2)+3\log\cos^2\theta)(n-a-b)!\nonumber\\
&\qquad\quad \left.+(6\pi\, l(u_\f^2)+\frac{3}{2}\log\cos^2\theta)(6\pi\, l(u_\ini^2)+\frac{3}{2}\log\cos^2\theta)(n-a-b-1)!\right)\left(1+\calo(n^{-1})\right)\label{grint8}
\end{align}
The common factor $(1+\calo(n^{-1}))$ is due to $1/n$ corrections to \eqref{asform1}, but these will be irrelevant when we sum the 4 types of diagrams and keep only the leading contribution. We already presented \eqref{grint5} in \eqref{1stres} but now have all other contributions listed as well. This allows us to finally analyze the growth of $t^{(n,2)}(\pv_\f,\pv_\ini)\propto \lambda^n\kappa^2$ by summing the contributions from the various diagrams. The leading growth goes like $(n-1)!$, we get contributions from \eqref{grint5}, (\ref{grint6}, \ref{grint7}) with $a=1$ and \eqref{grint8} with $a=b=1$, but in such a way that their sum cancels!  So instead we should look for growth of order $(n-2)!$. There are contributions from (\ref{grint6}, \ref{grint7}) with $a=2$ and \eqref{grint8} with $a=2, b=1$ and $a=1, b=2$, but also from the subleading terms in (\ref{grint6}, \ref{grint7}) with $a=1$ and \eqref{grint8} with $a=b=1$. Again their sum vanishes. Without being discouraged we investigate growth of the form $(n-3)!$. Now there are quite a few contributions: (\ref{grint6}, \ref{grint7}) with $a=3$ and \eqref{grint8} with $a=3, b=1$, $a=1, b=3$ and $a=2, b=2$, the subleading term of (\ref{grint6}, \ref{grint7}) with $a=2$ and \eqref{grint8} with $a=2, b=1$ and $a=1, b=2$ and also the subsubleading term of \eqref{grint8} with $a=b=1$. When we sum them again various cancellations happen but finally a non-zero contribution remains. The result is
\begin{equation}\boxed{
\ t^{(n,2)}(\pv_\f,\pv_\ini)\sim \frac{9}{2}(\cos\theta\,\log\cos^2\theta)^2 \kappa^2 \mu^{-\frac{3}{2}}\left(\frac{\lambda}{6\pi}\right)^n (n-3)!\ \label{finalren}}
\end{equation}
This formula for the asymptotic growth of the on-shell T-matrix of the model \eqref{model} is the key technical result of this paper. It establishes that non-relativistic 1-particle QM can exhibit a renormalon divergence in its perturbative series. In our derivation we saw that the $(n-1)!$ growth of \eqref{1stres} gets cancelled against diagrams with similar growth. As we remarked earlier this is as expected since \eqref{1stres} doesn't vanish at $\theta=0$ while the total result should, due to factorization and obvious absence of divergence at this value. Now observe that indeed the total result \eqref{finalren} vanishes at $\theta=0$, thus passing an important consistency check.

\subsection*{Borel summation: ambiguity and resolution}
The factorial growth \eqref{finalren} -- which for positive $\lambda$ is non sign-oscillating --  leads to a pole on the positive real axis of the Borel plane leading to an ambiguity in the Borel summation of $t^{(2)}$, the on-shell T-matrix at all order in $\lambda$ and second order in $\kappa$. By the formula \eqref{bamb} the ambiguity due to \eqref{finalren} is
\begin{eqnarray}
\mathrm{amb}\, t^{(2)}(\pv_\f,\pv_\ini)&=&\mp\frac{9\pi i}{2}(\cos\theta\,\log\cos^2\theta)^2 \kappa^2 \mu^{-\frac{3}{2}}e^{-\frac{6\pi}{\lambda}}\left(\frac{\lambda}{6\pi}\right)^2\\
&=&\mp 2\pi i(\cos\theta\,\log\cos^2\theta)^2 \kappa^2 \Lambda^{\frac{3}{2}}\left(\frac{\lambda}{4\pi}\right)^2\label{disc2}
\end{eqnarray}  
Given this ambiguity of the Borel summation procedure one needs to identify a physical principle to either decide the sign or cancel this extra imaginary part. 

To (re-)introduce this principle, let us revisit the terms that lead to the growth \eqref{finalren}. Let us collect those diagrams in (\ref{k2int5}-\ref{k2int8}) with integration of the $m$'th power of the logarithm. We can write their sum as follows
\begin{equation}
\tilde{t}_{m}^{(2)}(\pv_\f,\pv_\ini)=\int_{-\infty}^\infty dq\, f(q;\lambda;\pv_\f,\pv_\ini)\left(\frac{\lambda}{4\pi}\log \frac{q^2-p_\f^2}{\mu}\right)^{m}
\end{equation}
as we argued above such an integral grows like $(m-3)!$. Instead of performing the integrals and then summing over $m$ we could consider first summing and then integrating:
\begin{equation}
\tilde t^{(2)}(\pv_\f,\pv_\ini)=\int_{-\infty}^\infty dq \frac{f(q;\lambda;\pv_\f,\pv_\ini)}{1-\frac{\lambda}{4\pi}\log\frac{q^2-p_\f^2}{\mu}}\label{logsum}
\end{equation}
The divergence of the series of $\tilde t_m^{(2)}$ is now reflected in the divergence of the above integral. To make this a bit more explicit let us rewrite the integral above as\footnote{$f_\mathrm{e}(q)=\frac{1}{2}(f(q)+f(-q))$}
\begin{equation}
\tilde t^{(2)}(\pv_\f,\pv_\ini)=\int_{- p_\f^2}^{\infty} \frac{dE}{\sqrt{E+p_\f^2}} \frac{2f_\mathrm{e}(\sqrt{E+p_\f^2};\lambda;\pv_\f,\pv_\ini)}{1-\frac{\lambda}{4\pi}\log\frac{E}{\mu}}\end{equation}
The divergence is then due to the simple pole at $E=\Lambda$ and so can be avoided by moving it slightly below or above the real axis:
\begin{equation}
\tilde t_{\pm}^{(2)}(\pv_\f,\pv_\ini)=\int_{-p_\f^2}^\infty \frac{dE}{\sqrt{E+p_\f^2}} \frac{2f_\mathrm{e}(\sqrt{E+p_\f^2};\lambda;\pv_\f,\pv_\ini)}{1-\frac{\lambda}{4\pi}\log\frac{E\pm i\epsilon}{\mu}}\label{ieps}
\end{equation}
Of course this also introduces an ambiguity, which -- as we'll now discuss -- is the same as the ambiguity of the Borel summation. Apart from regularizing the integral as a principal value the $i\epsilon$ prescription in \eqref{ieps} also introduces an extra positive/negative imaginary part proportional to half the residue at $E=\Lambda$. This leads to
\begin{equation}
\mathrm{amb}\, \tilde t^{(2)}(\pv_\f,\pv_\ini)=\mp 2\pi i \Lambda \frac{f_\mathrm{e}(\sqrt{\Lambda+p_\f^2};\lambda;\pv_\f,\pv_\ini)}{\sqrt{\Lambda+p_\f^2}}\label{famb}
\end{equation}
In the limit $\lambda\rightarrow 0^+$ the renormalization invariant scale grows large, $\Lambda \rightarrow \infty$, and the ambiguity  \eqref{famb} is fully determined by the large $q$, small $\lambda$ behaviour of $f_\mathrm{e}(q;\lambda;\pv_\f,\pv_\ini)$. Using (\ref{largeqI1}, \ref{largeqI2}) and accounting for various cancellations -- identical to those observed previously -- the result is
\begin{equation}
f_\mathrm{e}(q;\lambda;\pv_\f,\pv_\ini)\sim \frac{\kappa^2}{2\pi}\left(\cos\theta\log\cos^2\theta\right)^2 q^{-4}(\log q^2)^{-2}
\end{equation}
Combining this expression with \eqref{famb} reproduces the Borel ambiguity \eqref{disc2} and shows explicitly that Borel summation with a prescription for the contour is in this case equivalent to a momentum integral with $i\epsilon$ prescription. 

The key point is that the $i\epsilon$ regularization introduced above is really that of Feynman. The physical choice -- which corresponds to the correct choice of ingoing-outgoing scattering boundary conditions -- is $p_\f^2+i\epsilon$ in the propagator and translates to $-i\epsilon$ in \eqref{ieps}, since $E=q^2-p_\f^2$. Although we reintroduced $i\epsilon$ in \eqref{ieps} it was in some sense always there, in that if we would have kept the $i\epsilon$ of our original Feynman rules \eqref{FR1} it would have appeared just like in \eqref{ieps} with the minus choice. Although at a given order it might have seemed to be perfectly valid to take the limit $\epsilon\rightarrow 0$ since this provided a sensible finite answer, we now see that this is more subtle and actually causes the renormalon pole to be on the positive real axis. In other words this limit does not commute with summation of the series:
\begin{equation}
\lim_{\epsilon\rightarrow 0} \int dq \sum_{n} a_n(q,\epsilon)\lambda^n\neq \sum_n \lim_{\epsilon\rightarrow 0} \int dq\, a_n(q,\epsilon) \lambda^n
\end{equation}
The left hand side provides a finite answer while the right hand side is a factorially diverging series. One can equivalently recover the finite answer on the left from the diverging series on the right by Borel summation, where the prescription in the Borel plane corresponding to the physical choice $-i\epsilon$ in \eqref{ieps} is to integrate along a contour that deforms the real axis {\it below} the renormalon pole, selecting the $+$ sign in \eqref{disc2}.

This observation indicates that renormalons of the perturbative on-shell T-matrix lead to an extra imaginary non-perturbative contribution to this on-shell T-matrix, that will not get canceled by additional non-perturbative corrections, but whose presence on the contrary is required by causality, i.e. outgoing waves only after scattering. In the next section we will recalculate $t^{(2)}(\pv_\f,\pv_\ini)$, exact and fully non-perturbatively in $\lambda$. As we will see this reproduces the results discussed here and also highlights more directly the role of the $i\epsilon$ prescription.

\section{Rederivation using exact Green's operator}\label{exact} 
In this section we will recompute $t^{(2)}(\pv_\f,\pv_\ini)$, the part of the on-shell T-matrix quadratic in $\kappa$, but now using operator formalism to do this exactly in $\lambda$. We first shortly review the relation between the operator formalism and the Born series in the standard perturbative setting and then point out how this can be easily adapted to find a series for the S-matrix which is perturbative in $\kappa$ but exact in $\lambda$. The key step is replacing the free Green's operator by the Green's operator of the 2d $\delta$ potential, which can be computed exactly.

\subsection*{Operator formalism and the Born series}
We start by reminding\footnote{See e.g. \cite{Taylor:72} for scattering theory in QM} the reader of the relation between the on-shell T-matrix $t$ and the off-shell T-operator $T$:
\begin{equation}
t(\pv_\f,\pv_\ini)=\langle \pv_\f|T(p_\f^2+i\epsilon)|\pv_\ini\rangle\label{tT}
\end{equation}
The off-shell T-operator -- defined for an arbitrary complex number $z$ not on the positive real axis -- is in turn determined in terms of the Green's operator/resolvent $G(z)$ and the potential $V=H-p^2$:
\begin{equation}
T(z)=V+VG(z)V\qquad\qquad G(z)=(z-H)^{-1}\label{Tdef}
\end{equation}
To connect to the standard perturbative Born series --  used in the previous sections -- one first rewrites  the Green's function for the interacting Hamiltonian in terms of that of the free Hamiltonian:
\begin{equation}
G(z)=(1-G_0(z)V)^{-1}G_0\qquad\qquad G_0=(z-p^2)^{-1}\label{gfs}
\end{equation}
It follows that
\begin{equation}
T(z)=V(1-G_0(z)V)^{-1}
\end{equation}
Inserting this expression in \eqref{tT} and expanding the inverse as a geometric series then yields the Born-series:
\begin{equation}
t(\pv_\f,\pv_\ini)=\sum_{n=0}^\infty\langle \pv_\f|V \left(G_0(p_\f^2+i\epsilon)V\right)^n|\pv_\ini\rangle
\end{equation}

\subsection*{Operator formalism and the $\lambda$-exact series}
Let us now consider our model \eqref{model}, where\footnote{For notational simplicity we absorb in this subsection the coupling constants $\lambda_0$ and $\kappa$ into $V_\star$ and $V_*$ respectively.} $V=V_\star+V_*$. Because we know the exact Green's operator $G_\star$ of the 2d $\delta$ Hamiltonian -- see below -- we might consider expressing the Green's function of the total Hamiltonian in terms of $G_\star$ and $V_*$ rather than $G_0$ and $V$:
\begin{equation}
G(z)=(1-G_\star(z)V_*)^{-1}G_\star\qquad\qquad G_\star(z) =(z-p^2-V_\star)^{-1}
\end{equation}
The expression for the T-operator obtained by inserting this formula in the definition \eqref{Tdef} is a bit more involved:
\begin{align}
T(z)=&T_\star(z)+V_*(1-G_\star (z)V_*)^{-1}
+V_\star(1-G_\star(z)V_*)^{-1} G_\star(z)V_*\nonumber\\
&+(1-V_*G_\star(z))^{-1}V_*\tilde G_\star (z)V_\star+V_\star(1-G_\star(z)V_*)^{-1} G_\star(z)V_* G_\star V_\star
\end{align}
where $T_\star(z)$ is the off-shell T-operator of the 2d $\delta$ potential.
As before we can now expand the inverses as geometric series, with the important and crucial difference that now this will give an expansion only in $V_*$, i.e. $\kappa$, while being exact in $\lambda$. The result is
\begin{align}
T(z)=&T_\star (z)+V_*\sum_{n=0}^\infty \left(G_\star(z) V_*\right)^n+V_\star\sum_{n=1}^\infty \left(G_\star(z) V_*\right)^n\nonumber\\
&+\sum_{n=1}^\infty \left(V_* G_\star(z)\right)^n V_\star+V_\star\sum_{n=1}^\infty \left(G_\star(z) V_*\right)^nG_\star(z) V_\star
\end{align}
The above might be more clear when expressed in diagrammatic language.  Apart from $T_\star$ there is a contribution from each diagram made out of an arbitrary number of vertices connected by propagators $\sim$ representing $G_\star$, with each vertex being a $*$, except the first or last vertex, which can also be $\star$. One has the following set of diagrams:
\begin{eqnarray}
*\sim\ldots\sim*\qquad \star\sim*\sim\ldots\sim*\qquad *\sim\ldots\sim*\sim\star \qquad \star\sim*\sim\ldots\sim*\sim\star\label{newdiag}
\end{eqnarray}
We stress again that this is a calculation perturbative in $\kappa$ while being exact in $\lambda$. By further expanding $G_\star$ in terms of $V_\star$ and $G_0$ one recovers the double expansion of the previous sections. The expansion of $G_\star$ has the diagramatic form
\begin{equation}
\sim\ \ =\ \ -\ \ +\ \ -\star-\ \ +\ \ -\star-\star-\ \ +\ \ \ldots
\end{equation}
The above -- computed via the renormalized Feynmann rules (\ref{FR1}, \ref{renrules}) -- is a series that converges to 
\begin{equation}
\langle \pv_1| G_\star (z)|\pv_2\rangle=\frac{(2\pi)^3\delta^3(\pv_{1}-\pv_{2})}{(z-p_2^2)}+2\pi\frac{\delta(q_1-q_2)}{(z-p_1^2)(z-p_2^2)}t_\star(z-q_2^2)\label{Gstar}
\end{equation}
where we refer to \eqref{tstar} for the definition of $t_\star$. That this is indeed the exact Greens function of the 2d $\delta$ model can be checked by comparing to results obtained through the non-perturbative definition of that model through self-adjoint extension.

We have now collected all ingredients to work out an alternative perturbation theory in $\kappa$, which is exact in $\lambda$. It consists of the diagrams \eqref{newdiag} with the Feynmann rules
\begin{align}
\star&\ :\ 2\pi\lambda\delta(q_{k-1}-q_k)\qquad\qquad\qquad\qquad *\ : \ \kappa \hat V_*(\pv_{k-1}-\pv_k)\label{FRel}\\
\sim&\ :\ \int \frac{d^3\pv_k}{(2\pi)^3}\langle \pv_{k-1}| G_\star (p_\f^2+i\epsilon)|\pv_k\rangle\label{exactG}
\end{align}
Using the above rules one readily computes the four diagrams of order $\kappa^2$:
\begin{eqnarray}
*\sim*&=&\int \frac{dq}{2\pi}\left(I^{(2,1)}(\pv_\f,\pv_\ini,q)+I^{(1,1)}(\pv_\f,q )I^{(1,1)}(\pv_\ini,q)^*t_{\star}(p_\f^2-q^2+i\epsilon)\right)\nonumber\\
\star\sim*\sim*&=&t_{\star}(u_\f^2)\int\frac{dq}{2\pi}\left(I^{(2,2)}(q_\f,\pv_\ini,q)+I^{(1,2)}(q_\f,q)I^{(1,1)}(\pv_\ini,q)^*t_{\star}(p_\f^2-q^2+i\epsilon)\right)\label{exactdiag}\\
*\sim*\sim\star&=&t_{\star}(u_\ini^2)\int\frac{dq}{2\pi}\left(I^{(2,2)}(q_\ini,\pv_\f,q)^*+I^{(1,2)}(q,q_\ini)I^{(1,1)}(\pv_\f,q)t_{\star}(p_\f^2-q^2+i\epsilon)\right)\nonumber\\
\star\sim*\sim*\sim\star&=&t_{\star}(u_\f^2)t_{\star}(u_\ini^2)\int\frac{dq}{2\pi}\left(I^{(2,3)}(q_\f,q_\ini,q)+I^{(1,2)}(q_\f,q)I^{(1,2)}(q,q_\ini;z)t_{\star}(p_\f^2-q^2+i\epsilon)\right)\nonumber
\end{eqnarray}
Let us now focus on the parts of the above result containing an integral over $t_\star$. Collecting the four contributions we can write them as 
\begin{equation}
\tilde t^{(2)}(\pv_\f,\pv_\ini)=\int_{-\infty}^\infty dq\, f(q;\lambda,\pv_\f,\pv_\ini)\frac{\lambda}{1-\frac{\lambda}{4\pi}\log\frac{q^2-p_\f^2-i\epsilon}{\mu}}
\end{equation}  
This reproduces (\ref{logsum}, \ref{ieps}) and confirms the resolution of the summation ambiguity by the $i\epsilon$ prescription discussed there, making it fully transparent via \eqref{tT} and \eqref{exactG}. Additionally it shows that there are no further non-perturbative effects that could cancel the extra imaginary contribution the $i\epsilon$ prescription introduces.

\section*{Acknowledgements}
We thank L. Akant, M. Unsal and especially T. Turgut for various discussions. CP and DVdB are partially supported by the Bo\u{g}azi\c{c}i University Research Fund under grant number 17B03P1.

\appendix

\section{Notation and conventions}\label{not}
All calculations in this paper deal with 1-particle quantum mechanics in three spatial dimensions, so we can work in units where $\hbar=2m=1$. This implies the only dimension remaining is length $\mathrm{L}$ and that $[E]=[p^2]=\mathrm{L}^{-2}$. The dimensions of some objects appearing in the main text are\footnote{Here $t$ is the on-shell T-matrix, not time.}
\begin{equation}
[S]=\mathrm{L}^3\,,\quad [t]=\mathrm{L}\,,\quad [\lambda_0]=[\lambda]=1\,,\quad[\kappa]=\mathrm{L}^{-1}\,,\quad [\mu]=[\Omega]=\mathrm{L}^{-2} 
\end{equation}

Most of the paper we work in momentum space, which is described by the vectors $\pv=(\uv,q)=(v,w,q)\in\mathbb{R}^3$ of which we'll denote the lengths by $p^2=\pv\cdot\pv$ and $u^2=\uv\cdot\uv$. We will only occasionally refer to position space, where we use the vectors ${\bf x}=(x,y,z)$. Note that sometimes $z$ will instead refer to an arbitray complex parameter. Our conventions for the Fourier transform are
\begin{equation}
\langle{\bf x}|\pv\rangle=e^{i {\bf x}\cdot\pv}\qquad \langle{\bf x}|{\bf x}'\rangle=\delta^3({\bf x-x'})\qquad \langle{\pv}|{\pv }'\rangle=(2\pi)^3\delta^3({\pv-\pv'})
\end{equation}
so that
\begin{equation}
\hat f(\pv)=\int d^3{\bf x}\, e^{-i\pv\cdot {\bf x}}f({\bf{x}})\qquad  f({\bf x})=\int \frac{d^3\pv}{(2\pi)^3}\, e^{i\pv\cdot {\bf x}}\hat f(\pv)
\end{equation} 

\section{Asymptotics of a key integral.}\label{intap}
For large $n$ there is the following asymptotic formula
\begin{equation}
I=\int_{-\infty}^\infty dq\, f(q)\left(\log(q^2-q_0)+a\right)^n\sim e^{\rho a}\rho^{-(n+1)}\,\left(\frac{n}{\rho}\right)^\sigma  \,n! \label{keyint}
\end{equation}
where the parameters $\rho$ and $\sigma$ are determined by the large $|q|$ behaviour of $f$, $q_0$ a positive constant and $a$ a complex number. More precisely the formula above is valid when $f$ decays for large $|q|$ as
\begin{equation}
f(q)\sim |q|^{-2\rho-1}\left(\log q^2\right)^\sigma\label{appdecay}
\end{equation}
Let us sketch how this formula is derived and why only the large $|q|$ region contributes while the contributions around $q^2\approx q_0^2$ cancel.

One starts by rewriting the integral as an integral over the even part of $f$ and splitting it over the regions $[0,q_0]$, $[q_0,\sqrt{2}q_0]$, $[\sqrt{2}q_0,\infty]$. In the first two regions one changes integration variables as $q=q_0\sqrt{1+e^{-t}}$ while in the third region $q=q_0\sqrt{1+e^{t}}$, so that $I=I_1+I_2+I_3$ with
\begin{eqnarray}
I_1&=&-\int_{0-i\pi}^{\infty-i\pi} dt \, e^{-t}\,g(t)\, (\tilde a-t)^n \quad\qquad\qquad\tilde a=a+\log q_0^2\\
I_2&=&\int_{0}^\infty dt \, e^{-t}\,g(t)\, (\tilde a-t)^n\ \qquad\qquad\qquad g(t)=\frac{q_0 f_\mathrm{e}(q_0\sqrt{1+e^{-t}})}{\sqrt{1+e^{-t}}}\\
I_3&=&\int_{0}^\infty dt \, e^{t}\,h(t)\, (\tilde a+t)^n\ \ \, \qquad\qquad\qquad h(t)=\frac{q_0 f_\mathrm{e}(q_0\sqrt{1+e^{t}})}{\sqrt{1+e^{t}}}
\end{eqnarray}
where $f_\mathrm{e}=\frac{1}{2}(f(q)+f(-q))$ is the even part of $f$. Each of these integrals will at large $n$ be dominated by the large $t$ region. Note that the first two integrals are very similar, except that the first is over a complex contour parallel to the real axis. Although each of this integrals grows like $n!$ these contributions cancel each other. One way to see this is that both integrals together are equal to a closed contour integral -- up to two vertical pieces which can be estimated not to grow factorially -- and the resulting residue contributions will only grow only with a powerlaw in $n$.  

This leaves us with the third contribution, using the assumed decay \eqref{appdecay} for $f$ one reproduces \eqref{keyint} by standard saddle point evaluation:
\begin{equation}
I_3\sim q_0^{-2\rho}\int^\infty dt\, e^{-\rho t}\,t^\sigma (\tilde a+t)^n\sim e^{\rho a} \,\rho^{-(n+1)}\left(\frac{n}{\rho}\right)^\sigma\, n!
\end{equation} 

  \bibliographystyle{JHEP}
      \bibliography{renormalon}
\end{document}